\documentclass[twocolumn,showpacs,preprintnumbers,amsmath,amssymb]{revtex4}
\newcommand{\be}{\begin{equation}}
\newcommand{\ee}{\end{equation}}
\newcommand{\bea}{\begin{eqnarray}}
\newcommand{\eea}{\end{eqnarray}}

\newcommand{\nn}{\nonumber}
\usepackage{graphicx}% Include figure files
\usepackage{dcolumn}% Align table columns on decimal point
\usepackage{bm}% bold math
\input epsf.sty

\begin{document}

\preprint{APS/123-QED}

\title{ Vlasov tokamak equilibria with shearad toroidal flow and anisotropic pressure
\footnote{Published in Phys. Plasmas {\bf 22}, 082505 (2015)}
\\
}% Force line breaks with \\

% Force line breaks with \\

\author{Ap Kuiroukidis$^{1}$, G. N. Throumoulopoulos$^{2}$, and H. Tasso$^{3}$}
 \altaffiliation[E-mail: kouirouki@astro.auth.gr,$\; \; $gthroum@uoi.gr,  $\; \;$  het@ipp.mpg.de]{}
\affiliation{
$^{1}$ Technological Education Institute of Serres, 62124 Serres, Greece\\
$^{2}$ University of Ioannina, Department of  Physics,
 GR 451 10 Ioannina, Greece \\
$^{3}$  Max-Planck-Institut
 f\"{u}r Plasmaphysik, D-85748 Garching, Germany
}%

%\affiliation{

\date{\today}% It is always \today, today,

%%%%%%%%%%%%%%%%%%%%%%%%%%%%%%%%%%%%%%%%%%%%%%%%%%%%%%%%%%%%%%%%%%%%%%%%%%%%%%%%%%%%%%%%%%%%%%%%%%%%%%%%%%
%%%%%%%%%%%%%%%%%%%%%%%%%%%%%%%%%%%%%%%%%%%%%%%%%%%%%%%%%%%%%%%%%%%%%%%%%%%%%%%%%%%%%%%%%%%%%%%%%%%%%%%%%%

\begin{abstract}

By choosing appropriate deformed Maxwellian ion and electron distribution functions depending on the two particle constants of motion, i.e. the energy and  toroidal angular momentum, we reduce the Vlasov axisymmetric equilibrium problem for quasineutral plasmas to a transcendental  Grad-Shafranov-like  equation. This equation is then solved numerically under the Dirichlet boundary condition for  an analytically prescribed boundary possessing a lower X-point to construct tokamak equilibria with toroidal sheared ion flow and  anisotropic pressure. Depending on the deformation of the  distribution functions these steady states can have   toroidal current densities  either peaked on the magnetic axis or hollow. These two kinds of equilibria may be regarded as  a bifurcation in connection with symmetry properties of the distribution functions on the magnetic axis.

%Valid PACS numbers may be entered using the \verb+\pacs{#1}+ command.
\end{abstract}

\maketitle

%%%%%%%%%%%%%%%%%%%%%%%%%%%%%%%%%%%%%%%%%%%%%%%%%%%%%%%%%%%%%%%%%%%%%%%%%%%%%%%%%%%%%%%%%%%%%%%%%%%%%%%%%%%%
%%%%%%%%%%%%%%%%%%%%%%%%%%%%%%%%%%%%%%%%%%%%%%%%%%%%%%%%%%%%%%%%%%%%%%%%%%%%%%%%%%%%%%%%%%%%%%%%%%%%%%%%%%%%

\section{Introduction}

Kinetic equilibria may provide broader and more precise
information than multifluid or MHD equilibria  for
axisymmetric plasmas governed by the well known Grad-Shafranov
(GS) equation. Because solving self-consistently the kinetic
equations is difficult, particularly in complicated geometries,
the majority of kinetic equilibrium solutions have been restricted
to  one-dimensional configurations in plane geometry
\cite{chan}-\cite{myni}. Of particular interest are equilibria
with sheared flows related to electric fields which play a role
in the transition to improved confinement regimes of magnetically
confined plasmas as the L-H transition and the formation of
Internal Transport Barriers. Construction of kinetic equilibria
is crucially related to the particle constants of motion, upon
which the distribution function depends. In the framework of
Maxwell-Vlasov theory only two constants of motion are known
for symmetric two-dimensional equilibria, i.e. the energy, $E$,
and the momentum $C_{x3}$ conjugate to the ignorable coordinate
$x_{3}$. Therefore for distribution functions of the form
$f(E,C_{x3})$, only macroscopic flows and currents along the
direction associated with $x_{3}$ can be derived, e.g., toroidal
flows for axisymmetric plasmas. The creation of poloidal flows
requiring additional constant(s) of motion remains an open
question.

In the presence of magnetic fields kinetic equilibria constructed
in the literature usually concern neutral non-flowing plasmas
in connection with the choice of functionally identical electron and
ion distribution functions in addition to the quasineutrality
condition implying vanishing electric fields \cite{chan}-\cite{belm}.
On physical grounds the assumption is oversimplifying because,
in addition to the aforementioned importance of electric fields,
it ignores the mass difference of ions and electrons. A more
realistic treatment permitting finite electric fields consists
in using only the quasineutrality to express the electrostatic
potential, $\Phi$, in terms of the components of the vector
potential involved \cite{myni}-\cite{thro}, e.g. for
two-dimensional equilibria in plane geometry, $\Phi(x,y)$ can
be expressed in terms of $A_{z}(x,y)$. In connection with the
present study we also refer to a previous paper \cite{tass},
in which it has been proved that the current on the magnetic
axis of an axisymmetric Vlasov equilibrium vanishes if the
gradient of the distribution function and the electric filed
are taken equal to zero on that axis. However, for the translational
symmetric two-dimensional configurations, quasineutral
equilibria with non-vanishing current density were explicitly
found in references \cite{schi}-\cite{thro}.

The aim of the present contribution is to construct  Vlasov
quasineutral  equilibria in toroidal axisymmetric  geometry. The major new
ingredient compared with plane equilibria is toroidicity
which plays an important role in tokamaks. A first step to the  construction of such equilibria was made in \cite{tath2014} for deformed Maxwellians  leading to GS-like equations describing either static equilibria or equilibria with rigid toroidal flow. However, complete construction of specific equilibrium configurations by solving the GS-like equations  was not made in \cite{tath2014}. Here we select  another form of exponentially deformed  Maxwellians  with exponent depending quadratically on the toroidal angular momentum. This choice   amenable to  analytic integrations in the velocity space results to more  pertinent tokamak equilibria with sheared toroidal flows.  The distribution functions chosen have  finite gradients
on the magnetic axis and provision is made so that  the  electric field vanishes  thereon. This is important because otherwise the resulting $\bf E\times \bf B$ drift on axis would create unphysical perpendicular flows.  Also, depending on the symmetry properties of the distribution functions on the magnetic axis determined by appropriate free parameters, derivation of equilibria with toroidal current densities either peaked on axis or hollow is possible.  It is shown that
the above described  procedure leads to a GS-like equation with a transcendental
right-hand side which  is then solved numerically for a fixed diverted
boundary. Up to the best of the authors knowledge such  Vlasov tokamak  equilibria in toroidal geometry are constructed  for the first time.

The organization of the paper is as follows:
In Sec.  II we present the general setting of the equilibrium
equations in the framework of Vlasov theory. In Sec.  III making the aforementioned choice of  distribution functions we derive  the GS-like equation. In Sec.  IV  we examine the equilibrium properties by calculating  macroscopic  quantities  as the toroidal current density, fluid velocities  and  pressure.   Finally the main conclusions are summarized in Sec.  V.
\\

%%%%%%%%%%%%%%%%%%%%%%%%%%%%%%%%%%%%%%%%%%%%%%%%%%%%%%%%%%%%%%%%%%%%%%%%%%%%%%%%%%%%%%%%%%%%%%%%%%
%%%%%%%%%%%%%%%%%%%%%%%%%%%%%%%%%%%%%%%%%%%%%%%%%%%%%%%%%%%%%%%%%%%%%%%%%%%%%%%%%%%%%%%%%%%%%%%%%%
%%%%%%%%%%%%%%%%%%%%%%%%%%%%%%%%%%%%%%%%%%%%%%%%%%%%%%%%%%%%%%%%%%%%%%%%%%%%%%%%%%%%%%%%%%%%%%%%%
\section{General Development of the Kinetic Framework}

We consider axisymmetric toroidal plasmas and employ cylindrical
coordinates $(z,R,\phi)$, with z corresponding to the axis of symmetry
and $\phi$ being the toroidal, ignorable angle. The coordinate
system is illustrated in Figure 6.1 of reference \cite{frei}.
Axisymmetry means that any quantity depends solely on $z$ and $R$.
The toroidicity relates to the scale factor $1/R$ appearing in the
various equations, e.g. equation (\ref{magn}) for the magnetic field; also
the differential operator in connection with Ampere's law is the
elliptic operator on the LHS of Eq. (\ref{gs1}) in Sec.  III while it is the Laplace
operator in plane geometry. As already mentioned in Sec.  I in
axisymmetric plasmas only two constants of motion are known,
the energy $E$ and the angular momentum $C$. Using convenient units
by setting the magnetic permeability of vacuum equal to unity,
we have for the ions
\bea
\label{ioncs}
E_{i}&=&e\Phi(R,z)+\frac{M}{2}(v_{R}^{2}+v_{\phi}^{2}+v_{z}^{2})\nn
\\
C_{i}&=&MRv_{\phi}+eRA_{\phi}
\eea
while for the electrons we have
\bea
\label{elecs}
E_{e}&=&-e\Phi(R,z)+\frac{m}{2}(v_{R}^{2}+v_{\phi}^{2}+v_{z}^{2})\nn
\\
C_{e}&=&mRv_{\phi}-eRA_{\phi}
\eea
where $\Phi$ is the electrostatic potential, $A_{\phi}$ is the
toroidal component of the vector potential, $M$ and $m$ are
the masses of ions and electrons respectively, $v_{R},\; v_{\phi},\; v_{z}$
are the components of the particle velocities along the basis
vectors of the cylindrical coordinate system. The charge $e$
is taken as the absolute value of the electron charge.

The solutions of the ion and electron Vlasov equations
are given as
\bea
\label{vlasol}
f_{i}&=&f_{i}(E_{i},C_{i})\nn \\
f_{e}&=&f_{e}(E_{e},C_{e})
\eea
with a normalization of $f_{i},\; f_{e}$ equal to the total
number of particles $N$ so that the densities are given by
\bea
\label{dens}
n_{i}&=&\int f_{i}(E_{i},C_{i})d^{3}v\nn \\
n_{e}&=&\int f_{e}(E_{e},C_{e})d^{3}v
\eea
The electric current density is given by
\bea
\label{curr}
J_{\phi}=e\int v_{\phi}f_{i}(E_{i},C_{i})d^{3}v-
e\int v_{\phi}f_{e}(E_{e},C_{e})d^{3}v
\eea

We assume quasineutrality through the whole plasma instead of
Poisson equation for the electric potential. Also we demand
that the electric field, ${\bf E}$, vanishes on axis. Otherwise,
as already mentioned in Sec.  I,  the $R$ component of the electric field in combination with the
purely toroidal magnetic field on axis would lead to an
unphysical ${\bf E}\times{\bf B}$ drift parallel to the axis
of symmetry. A similar condition of vanishing ${\bf E}$ on axis
was adopted in  \cite{tass} for a near axis consideration
of the Vlasov equation. This leads to
\bea
\label{qsnet}
n_{i}=n_{e}
\eea
everywhere inside the plasma and in particular
\bea
\label{axnet}
\nabla n_{i}=\nabla n_{e}
\eea
on the magnetic axis. We introduce now $\Psi=RA_{\phi}$ as the
poloidal flux around the magnetic axis, a function which labels
the magnetic surfaces and can be taken equal to zero on the
magnetic axis. Also, we compute explicitly both sides of
Eqs.  (\ref{qsnet}) and (\ref{axnet}) using (\ref{dens})
to obtain
\bea
\label{equn}
\int f_{i}(E_{i},C_{i})d^{3}v=\int f_{e}(E_{e},C_{e})d^{3}v
\eea
and
\bea
\label{equdn}
\nabla n_{i}&=&\int \left[e\frac{\partial f_{i}}{\partial E_{i}}\nabla \Phi\right.
+e\frac{\partial f_{i}}{\partial C_{i}}\nabla \Psi +\nn \\
&+&\left.Mv_{\phi}\frac{\partial f_{i}}{\partial C_{i}}\nabla R\right]\nn
\\
\nabla n_{e}&=&\int \left[-e\frac{\partial f_{e}}{\partial E_{e}}\nabla \Phi\right.
-e\frac{\partial f_{e}}{\partial C_{e}}\nabla \Psi +\nn \\
&+&\left.mv_{\phi}\frac{\partial f_{e}}{\partial C_{e}}\nabla R\right]
\eea
We see from Eq. (\ref{equn}) that the electrostatic potential
will be, in general, a function of $\Psi$ and $R$ in contrast with
the scale factor free two-dimensional case treated in references
\cite{schi}-\cite{thro}. Similarly Eqs.  (\ref{equdn}) show that
the gradient of the electrostatic potential does not necessarily
vanish at the magnetic axis given by $\Psi=\nabla \Psi=\nabla\Phi=0$
because $\frac{\partial f_{i}}{\partial C_{i}}$ and
$\frac{\partial f_{e}}{\partial C_{e}}$ cannot vanish unless the toroidal
current vanishes also. This leads us to look for special choices
for $f_{i},\; f_{e}$ which would allow us to have in general $J_{\phi}$ different
from zero, while at the same time satisfy $\Psi=\nabla \Psi=\nabla\Phi=0$
on axis.

With this in mind we observe that in order to satisfy Eq.  (\ref{axnet})
at the magnetic axis, a sufficient condition would be that the equality of
the last terms of Eqs.  (\ref{equdn}) is satisfied at $\Psi=0$.
We thus choose
\bea
\label{distrs}
f_{i}&=&n_{0}\left(\frac{M\beta_{i}}{2\pi}\right)^{3/2}exp(-\beta_{i}E_{i})\times \nn
\\
&\times&\left[1+\alpha_{i}
exp\left(\beta_{i}^{2}\frac{V_{i\phi}^{2}}{R_{n}^{2}}(C_{i}+\tilde{A})^{2}\right)\right]\nn \\
f_{e}&=&n_{0}\left(\frac{m\beta_{e}}{2\pi}\right)^{3/2}exp(-\beta_{e}E_{e})\times \nn
\\
&\times&\left[1+\alpha_{e}
exp\left(\beta_{e}^{2}\frac{V_{e\phi}^{2}}{R_{n}^{2}}(C_{e}+\tilde{B})^{2}\right)\right]\nn \\
\eea
with $\beta_{s}=1/(kT_{s}),\; \; (s=i,e)$ where $E_{i},\; C_{i},\; E_{e},\; C_{e}$
are given by Eqs.  (\ref{ioncs}), (\ref{elecs}) and
$0<\alpha _{i},\; \alpha_{e}\leq 1$ are constants. Also $V_{i\phi},\; V_{e\phi},\; R_{n}$
are constants to be determined below. The free parameters $\tilde{A}$ and $\tilde{B}$ are crucial in determining the equilibrium characteristics. Specifically, for $\tilde{A}=\tilde{B}=0$ the distribution functions become symmetric with respect to $v_\phi$ on the magnetic axis (where $\Psi=0$) resulting in vanishing toroidal current densities thereon and therefore hollow current density profiles. In contrast, peaked current density profiles are derived for 
$\tilde{A}\neq 0$ or/and $\tilde{B}\neq 0$, a choice which breaks the $v_\phi$- symmetry of $f_i$ and $f_e$ on axis. To derive peaked  $J_\phi$ on axis and vanishing on the boundary as well as macroscopic ion flow we have 
chosen 
\be
\tilde{A}\beta_{i}\frac{V_{i\phi}}{R_{n}}:=A=-1\ \ \mbox{and} \ \ \tilde{B}\beta_{e}\frac{V_{e\phi}}{R_{n}}:=A=-1
                               \label{AB}
\ee
 Irrespective of the values of $\tilde{A}$ and $\tilde{B}$ the integrations in the velocity space can be performed analytically.  The characteristics of  both kinds of equilibria, peaked and hollow, will be presented in Sec.  IV. However, since for $\tilde{A}\neq 0$ and $\tilde{B} \neq 0$ certain expressions become  rather complicated, in the reminder of the report  complete  analytic results will be given only for $\tilde{A}=\tilde{B}=0$. 
%In the following analytic results will be given for the 
%in order to obtain a non-hollow current
%density profile in Fig. 4, below. We give below the analysis for the case
%$A=0$ which renders the hollow profiles and just present also the results
%for the case $A=-1$ which renders the non-hollow profiles.

In order now to satisfy Eq. 
(\ref{axnet}), as we have already stated,  we impose
equality of the last terms of Eqs.  (\ref{equdn}).
This can be accomplished if we choose the following conditions to be satisfied
\bea
\label{condts}
\beta_{i}MV_{i\phi}^{2}=\beta_{e}mV_{e\phi}^{2}
\Longrightarrow V_{e\phi}=\sqrt{\frac{M\beta_{i}}{m\beta_{e}}}V_{i\phi}\nn \\
\alpha_{e}=\sqrt{\frac{m\beta_{e}}{M\beta_{i}}}\alpha_{i}\nn \\
R_{n}^{2}>2\beta_{i}MV_{i\phi}^{2}R_{max}^{2}
\eea
where $R_{max}$ is the maximum value of $R$, to be specified below.
We set $\beta:=2\beta_{i}MV_{i\phi}^{2}=2\beta_{e}mV_{e\phi}^{2}$.

Using $f_{i},\; f_{e}$ in Eqs.  (\ref{dens})
and then  Eq. 
(\ref{qsnet}) we  obtain
\bea
\label{potent}
& &exp(-e(\beta_{i}+\beta_{e})\Phi)=F(R,\Psi):=\frac{F_{e}}{F_{i}}\nn \\
F_{e}&:=&1+\frac{\alpha_{e}}{\sqrt{1-\beta(R^{2}/R_{n}^{2})}}
exp\left(\frac{\beta_{e}^{2}V_{e\phi}^{2}e^{2}\Psi^{2}/R_{n}^{2}}{1-\beta(R^{2}/R_{n}^{2})}\right)\nn
\\
F_{i}&:=&1+\frac{\alpha_{i}}{\sqrt{1-\beta(R^{2}/R_{n}^{2})}}
exp\left(\frac{\beta_{i}^{2}V_{i\phi}^{2}e^{2}\Psi^{2}/R_{n}^{2}}{1-\beta(R^{2}/R_{n}^{2})}\right)\nn
\\
\eea
Thus we verify that in general $\Phi=\Phi(R,\Psi)$. From this we find that
\bea
\label{potent1}
exp(-e\beta_{i}\Phi)=F^{\frac{\beta_{i}}{\beta_{i}+\beta_{e}}}\nn \\
exp(e\beta_{e}\Phi)=F^{\frac{-\beta_{e}}{\beta_{i}+\beta_{e}}}
\eea
Computing the current density of Eq.  (\ref{curr}) we obtain
\bea
\label{curr1}
J_{\phi}&=&en_{0}\alpha_{i}exp(-e\beta_{i}\Phi)
\frac{2\beta_{i}V_{i\phi}^{2}eR\Psi/R_{n}^{2}}{[1-\beta(R^{2}/R_{n}^{2})]^{3/2}}\times\nn
\\
&\times &exp\left(\frac{\beta_{i}^{2}V_{i\phi}^{2}e^{2}\Psi^{2}/R_{n}^{2}}{1-\beta(R^{2}/R_{n}^{2})}\right)+\nn
\\
&+&en_{0}\alpha_{e}exp(e\beta_{e}\Phi)
\frac{2\beta_{e}V_{e\phi}^{2}eR\Psi/R_{n}^{2}}{[1-\beta(R^{2}/R_{n}^{2})]^{3/2}}\times\nn
\\
&\times &exp\left(\frac{\beta_{e}^{2}V_{e\phi}^{2}e^{2}\Psi^{2}/R_{n}^{2}}{1-\beta(R^{2}/R_{n}^{2})}\right)
\eea
where Eqs. (\ref{potent}) and (\ref{potent1}) are used.

The magnetic field can be written as
\bea
\label{magn}
{\bf B}=\frac{I_{0}}{R}{\bf e}_{\phi}+\nabla\Psi\times\frac{{\bf e}_{\phi}}{R}
\eea
where $I_{0}/R$ is the magnitude of a vacuum toroidal field at some value of $R$.
Projecting the curl of ${\bf B}$ on ${\bf e}_{\phi}$ and equating it with the
current density we obtain
the GS-like equation to be specified  and solved
numerically in  Section III.

%%%%%%%%%%%%%%%%%%%%%%%%%%%%%%%%%%%%%%%%%%%%%%%%%%%%%%%%%%%%%%%%%%%%%%%%%%%%%%%%%%%%%%%%
%%%%%%%%%%%%%%%%%%%%%%%%%%%%%%%%%%%%%%%%%%%%%%%%%%%%%%%%%%%%%%%%%%%%%%%%%%%%%%%%%%%%%%%%

\section{The Grad-Shafranov-like equation}\

For ITER-like equilibria we have $R_{0}=6.2m$ for the major radius and $a=2m$ the minor radius and therefore the aspect ratio is $\epsilon _{0}=0.32$. Thus the maximum distance perpendicular to the axis of symmetry is found to be $R_{max}=R_{0}(1+\epsilon_0)$. Introducing the  normalized
variables $\rho:=R/R_{0},\; \zeta:=z/R_{0}$ they range as $0.7\leq \rho\leq 1.2$
and $|\zeta |\leq 0.6$.
Assigning  to the temperatures the values $kT_{s}\simeq 10keV,\; (s=i,e)$,
we find  $\beta\simeq 0.021$. Thus we must have $R_{n}>R_{max}\sqrt{\beta}\simeq 1.2m$.
We choose $R_{n}=2m$ and define $\rho_{n}:=R_{n}/R_{0}$.
Also we take as typical values $n_{0}\simeq 10^{19}m^{-3}$ and
$V_{i\phi}\simeq 10^{5}m/sec$.

From  the restriction  $0<\alpha_{i}\leq 1$ adopted it follows
$\alpha_{e}\simeq 0.023\alpha_{i}$. We proceed now to a
normalization of all the above involved quantities.
In particular,  the poloidal flux is normalized
as $\Psi_{n}:=\Psi/\Psi_{0}$, where $\Psi_{0}:=R_{n}/(\beta_{i}V_{i\phi}e)\simeq 0.2 Wb$.
Then we have
\bea
\label{ef}
F=\frac{1+\frac{\alpha_{e}}{\sqrt{1-\beta(\rho^{2}/\rho_{n}^{2})}}
exp\left(\frac{\Psi_{n}^{2}}{1-\beta(\rho^{2}/\rho_{n}^{2})}\right)}
{1+\frac{\alpha_{i}}{\sqrt{1-\beta(\rho^{2}/\rho_{n}^{2})}}
exp\left(\frac{\Psi_{n}^{2}}{1-\beta(\rho^{2}/\rho_{n}^{2})}\right)}
\eea
with the form of  Eqs. (\ref{potent1}) remaining unaffected. Subsequently  we obtain
\bea
\label{curr2}
J_{\phi}&=&
[C_{i}\alpha_{i}exp(-e\beta_{i}\Phi)+C_{e}\alpha_{e}exp(e\beta_{e}\Phi)]\times\nn
\\
&\times&\frac{\beta(\rho^{2}/\rho_{n}^{2})\Psi_{n}}{[1-\beta(\rho^{2}/\rho_{n}^{2})]^{3/2}}
exp\left(\frac{\Psi_{n}^{2}}{1-\beta(\rho^{2}/\rho_{n}^{2})}\right)
\eea
The numerical values of the constants are $C_{i}\simeq 7405$ and $C_{e}=(M/m)C_{i}$.
The completely normalized GS-like equation turns out to be
\bea
\frac{\partial^{2}\Psi_{n}}{\partial\rho^{2}}-\frac{1}{\rho}
\frac{\partial\Psi_{n}}{\partial\rho}+
\frac{\partial^{2}\Psi_{n}}{\partial\zeta^{2}}=-J_{\phi}
                                                     \label{gs1}
\eea
where  $\Phi(R,\Psi_n)$ is determined by Eq. (\ref{potent}).
It is noted that (\ref{gs1}) holds for arbitrary values of $\tilde{A}$ and $\tilde{B}$. 

\section{Equilibria and equilibrium properties}\

First we define the fixed boundary coinciding with outermost magnetic surface as follows.   The equation for the upper part of the bounding surface,
which if taken to hold for the lower part as well would give a symmetric
boundary, is
\bea
\label{upbound}
\rho_{b}&=&1+\epsilon _{0}cos(\tau +\alpha sin(\tau))\nn \\
\zeta_{b}&=&\zeta_{max}sin(\tau)
\eea
where $\zeta_{max}=\kappa\epsilon _{0}$
with $\delta=(1-\rho_{\delta})/\epsilon_{0}$,
and $\alpha=sin^{-1}(\delta)$. Thus the following relations hold:
$\rho_{\delta}=1-\delta\epsilon_{0}$ and
$\theta_{\delta}=\pi-tan^{-1}(\kappa/\delta)$ (see Fig. 1)
The parameter $\tau$ is any increasing function of the polar angle
$\theta$, satisfying $\tau(0)=0$, $\tau(\pi)=\pi$ and $\tau(\theta_{\delta})=\pi/2$.
In our model we take
\bea
\label{taf}
\tau(\theta)&=&t_{0}\theta^{2}+t_{1}\theta^{n}\nn \\
t_{0}&=&\frac{\theta_{\delta}^{n}-\frac{1}{2}\pi^{n}}{\pi\theta_{\delta}^{n}-\theta_{\delta}^{2}\pi^{n-1}}
\nn \\
t_{1}&=&\frac{-\theta_{\delta}^{2}+\frac{1}{2}\pi^{2}}{\pi\theta_{\delta}^{n}-\theta_{\delta}^{2}\pi^{n-1}}
\eea
with $n=8$. In order to complete the asymmetric bounding curve we specify now
the lower part of it $(\zeta<0)$ as follows. The left lower  branch of the curve is given by
\bea
\label{left}
\rho_{b}&=&1+\epsilon_{0}cos(\theta)\nn \\
\zeta_{b}&=&-[2p_{1}\epsilon_{0}(1+cos\theta)]^{1/2}\nn \\
p_{1}&=&\frac{\zeta_{max}^{2}}{2\epsilon_{0}(1+cos\theta_{\delta})},\; \; \; \;
(\pi\leq \theta \leq 2\pi-\theta_{\delta})
\eea
while the right  lower branch of the curve is given by
\bea
\label{right}
\rho_{b}&=&1+\epsilon_{0}cos(\theta)\nn \\
\zeta_{b}&=&-[2p_{2}\epsilon_{0}(1-cos\theta)]^{1/2}\nn \\
p_{2}&=&\frac{\zeta_{max}^{2}}{2\epsilon_{0}(1-cos\theta_{\delta})},\; \; \; \;
(2\pi-\theta_{\delta}\leq \theta \leq 2\pi)
\eea
The divertor  X-point for the asymmetric equilibrium is
located at $\rho_{X}=1+\epsilon_{0}cos\theta_{\delta}=0.9139$
and $\zeta_{X}=-\zeta_{max}=-0.6105$.

\begin{figure}[ht!]
\centerline{\mbox {\epsfxsize=10.cm \epsfysize=8.cm \epsfbox{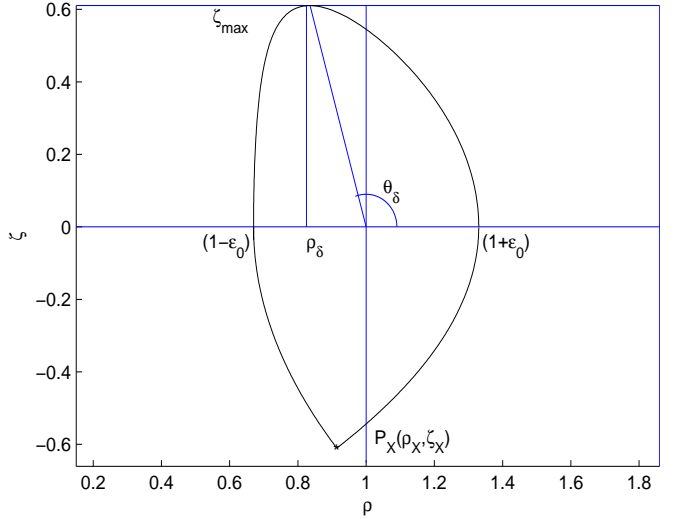}}}
\caption[]{The boundary determined by the parametric Eqs. (20)-(23)}
\label{fig1}
\end{figure}

\begin{figure}[ht!]
\centerline{\mbox {\epsfxsize=10.cm \epsfysize=8.cm \epsfbox{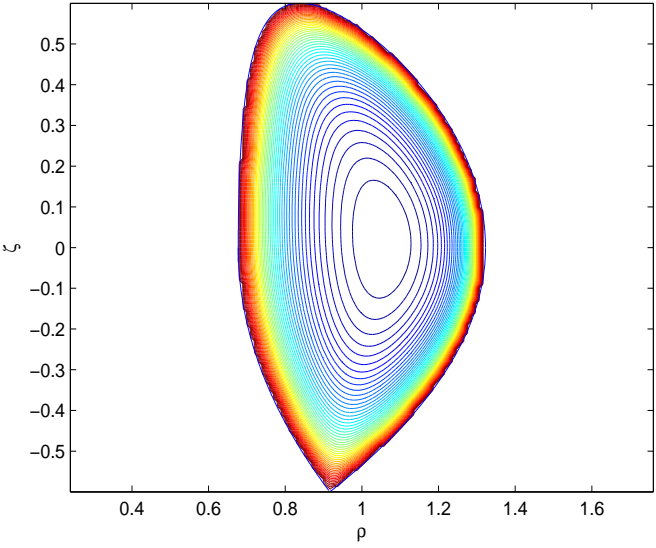}}}
\caption[]{The equilibrium configuration in connection with the numerical solution of the GS-like equation [Eqs. (17)-(18)] for peaked toroidal current density $A=B=-1$ [Eq. (\ref{AB})]. The magnetic surfaces are nearly the same for the respective equilibrium  with hollow $J_\phi$ ($\tilde{A}=\tilde{B}=0$).  }
\label{fig2}
\end{figure}

We adopt the numerical integration scheme described in detail in Sec.  4 of
our previous paper \cite{kui1} using the nine point formula. Using the   numerical algorithm associated with this formula  for peaked toroidal current density ($A=B=-1$) we
obtained the equilibrium shown in Fig. 2.   The magnetic axis was found by the
numerical procedure to be located at $(\rho_{a},\zeta_{a})=(1.0574,0.02)$.
The numerical procedure converged to the actual solution of Fig. 2 after
$N=2027$ iterations with a finite difference step size of $h=0.01$.
We have taken $\alpha_{i}=0.0015$. At the boundary we have set
$\Psi_{n}(b)=1.0$, while the numerical integration scheme turned out
to give for the magnetic axis $\Psi_{n}(a)=0.0$. The configuration of Fig. 2 remains nearly unaffected when $J_\phi$ changes to hollow ($\tilde{A}=\tilde{B}=0$).

%\begin{figure}[ht!]
%\centerline{\mbox {\epsfxsize=10.cm \epsfysize=8.cm \epsfbox{sxnma3.eps}}}
%\caption[]{The equilibrium described in the text.}
%\label{fig3}
%\end{figure}

%We have constructed four equilibria. The first is obtained
%for $X_{1}=X_{2}=0.5$, $P_{1}=P_{2}=0.013$, $\Phi_{1}=\Phi_{2}=3.384\times 10^{-4}$.
%The freely specified constants are given by $c_{6}=4.0$, $c_{5}=-0.5$,
%$c_{4}=5.0$, $c_{3}=-0.5$, $c_{2}=1.0$, $c_{1}=-0.1$ and $c_{0}=1.0$.
%The center of the equilibrium is at
%$(\rho_{c},\zeta_{c})=(1.04,0.02)$ and the X-point at $(\rho_{X},\zeta_{X})=(0.95,-0.51)$.
%At the center we have $u_{c}=-1.0325$ and at the boundary $u_{b}=-0.872$.
%The inverse aspect ratio is $\eps=0.127$, the elongation is $\kappa=3.85$
%and the triangularity is $\delta=0.51$.
%%%%%%%%%%%%%%%%%%%%%%%%%%%%%%%%%%%%%%%%%%%%%%%%%%%%%%%%%%%%%%%%%%%%%%%%%%%%%%%%%%%%%%
%\newpage

On the basis of the solution constructed we  examined the characteristics for   both equilibria with peaked and hollow $J_\phi$ by calculating the safety factor and certain equilibrium quantities as follows.   For axisymmetric plasmas the safety factor can be put in the form
%is given by \cite{wess}, \cite{whit}
\bea
\label{qfact}
q(\Psi_n)=\frac{1}{2\pi}\int_{0}^{2\pi}Q(\Psi_n,\theta)d\theta
\eea
where
\bea
\label{qfact1}
Q(\Psi_n,\theta)=Q_{0}
\frac{[(\rho-1)^{2}+\zeta^{2}]}{|(\rho-1)(\Psi_n)_{,\rho}+\zeta (\Psi_n)_{,\zeta}|}
\eea
and $Q_{0}\simeq 2.28$ is a dimensionless constant.
%Irrespective of the shape of $J_\phi$ the  safety factor increases monotonically from the magnetic axis to the plasma boundary. 
The variation of the  safety factor for peaked $J_\phi$ increasing monotonically from the magnetic axis to the plasma boundary is shown in Fig. 3. The profile of  $J_\phi$ on the midplane $\zeta=0$ which vanishes on the boundary is shown in 
%Fig. 4a,
Fig. 4(above),
while the respective hollow, current-hole like profile of $J_\phi$ is shown in %Fig. 4b. 
Fig. 4(below). Since in the later case $J_\phi$ vanishes on axis the safety factor tends to infinity thereon and the equilibrium in the central region has negative magnetic shear. Current hole equilibria of this kind have been observed in JET \cite{hast} and JT-60U \cite{fusu}.
% The latter equilibrium is exceptional because tokamak equilibria with hollow current densities are usually associated with reversed magnetic shear. Another    nonlinear magnetohydrodynamic equilibrium with monotonically increasing $q$ but hollow $J_\phi$  was found  in  \cite{kuth2014a}.
%The toroidal current density is plotted
%in Fig. 5.

\begin{figure}[ht!]
\centerline{\mbox {\epsfxsize=10.cm \epsfysize=8.cm \epsfbox{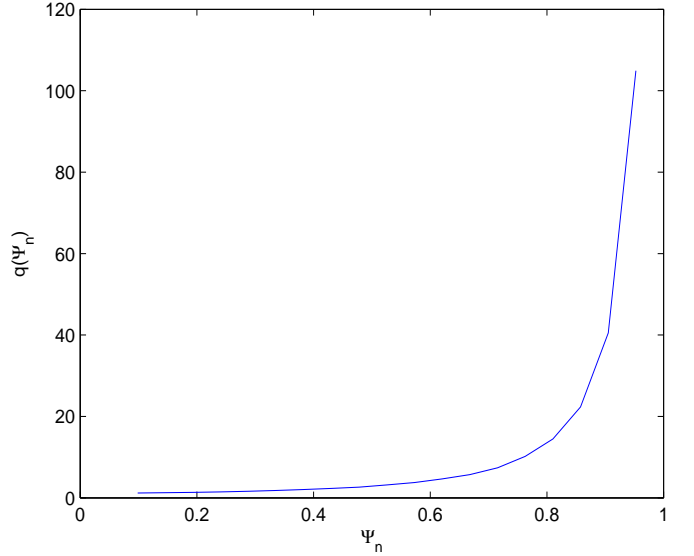}}}
\caption[]{Variation of the safety factor for the equilibrium of Fig. 2 from the magnetic axis to the boundary for peaked on axis $J_\phi$. At the magnetic axis
we have $q_{a}=1.17$. }
%The shape of the respective curve of the safety factor for the equilibrium with hollow $J_\phi$ is nearly the same.}
\label{fig3}
\end{figure}

\begin{figure}[ht!]
\centerline{\mbox {\epsfxsize=10.cm \epsfysize=8.cm \epsfbox{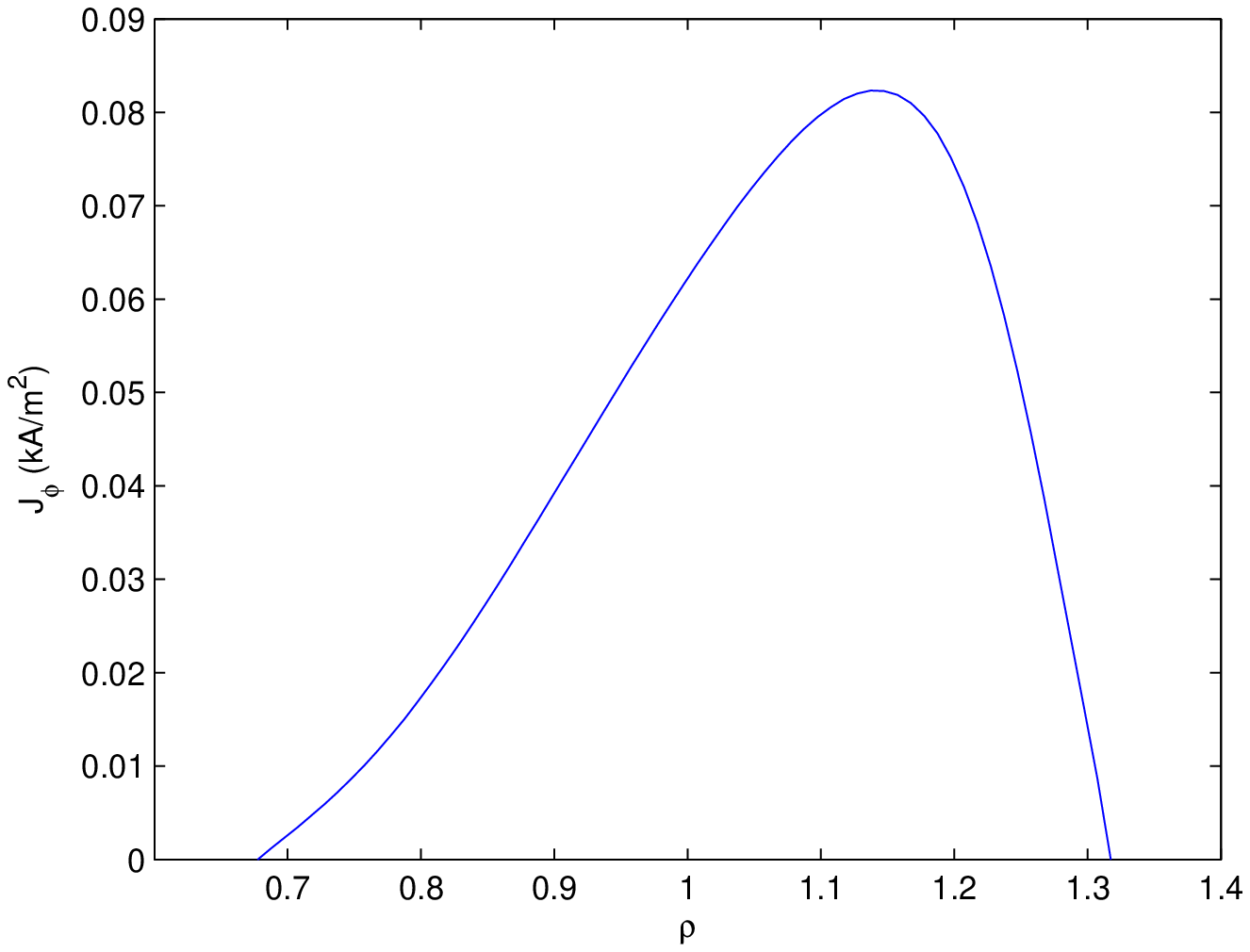}}}
\centerline{\mbox {\epsfxsize=10.cm \epsfysize=8.cm \epsfbox{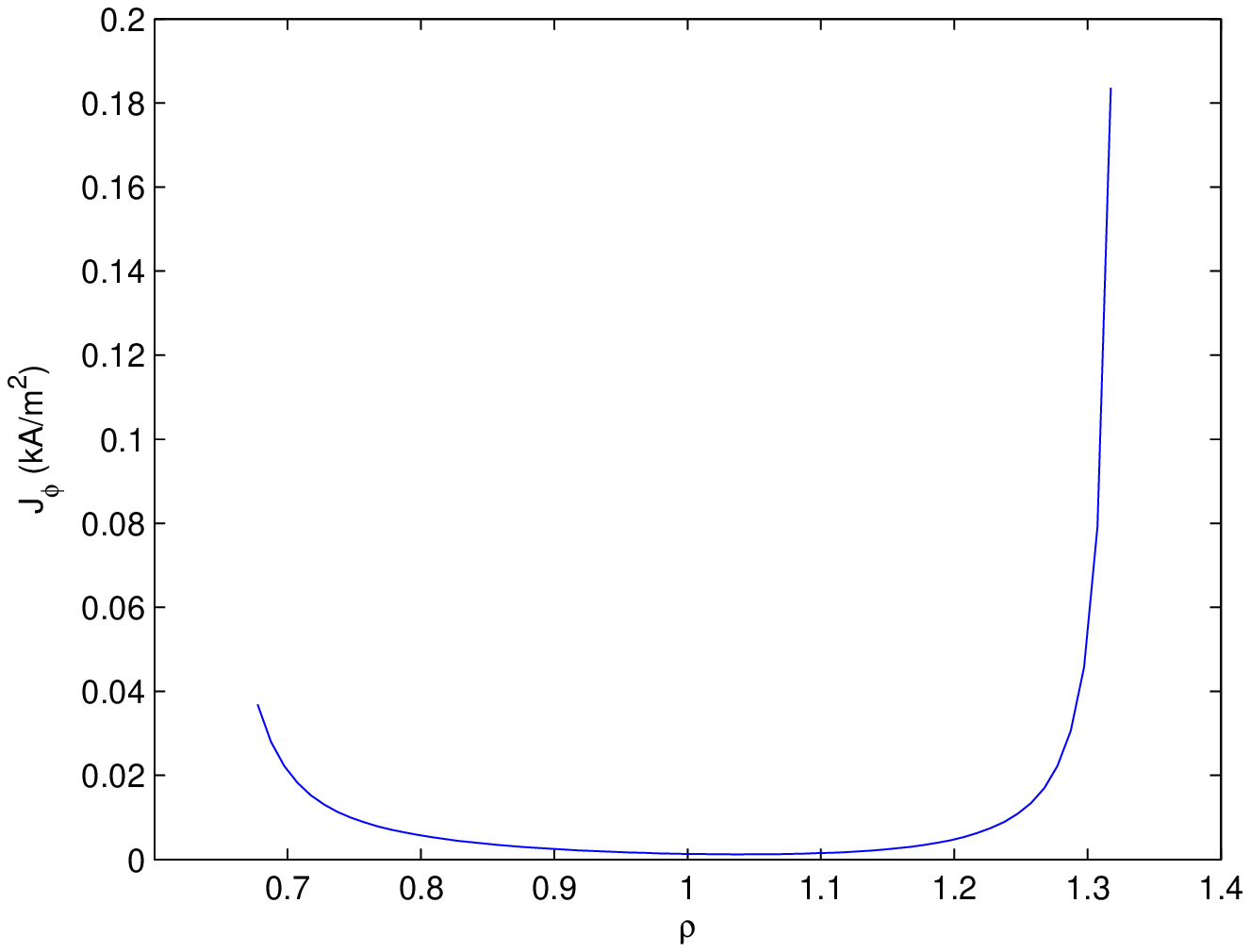}}}
\caption[]{Peaked on axis toroidal current density for $A=B=-1$ in connection with the equilibrium of Fig. 2. 
%(Fig. 4a)
(above)
 and respective hollow $J_\phi$ for $\tilde{A}=\tilde{B}=0$ 
%(Fig. 4b).
(below).}
\label{fig4}
\end{figure}

The toroidal ion fluid velocity $U_{i\phi}$ is given by
\bea
\label{uiphi}
U_{i\phi}&=&\frac{1}{n} \int \, v_\phi f d^3 v \nn \\
&=&
\frac{
\frac{V_{i\phi}\alpha_{i}(\rho/\rho_{n})\Psi_{n}}{[1-\beta(\rho^{2}/\rho_{n}^{2})]^{3/2}}
exp\left(\frac{\Psi_{n}^{2}}{1-\beta(\rho^{2}/\rho_{n}^{2})}\right)}
{1+\frac{\alpha_{i}}{\sqrt{1-\beta(\rho^{2}/\rho_{n}^{2})}}
exp\left(\frac{\Psi_{n}^{2}}{1-\beta(\rho^{2}/\rho_{n}^{2})}\right)}
\eea
This  for the two equilibria considered is  plotted in Fig. 5. Note that the   profile shapes are similar to the respective shapes of the current density profiles. Associated electric  field profiles are given in Fig. 6. 
\begin{figure}[ht!]
\centerline{\mbox {\epsfxsize=10.cm \epsfysize=8.cm \epsfbox{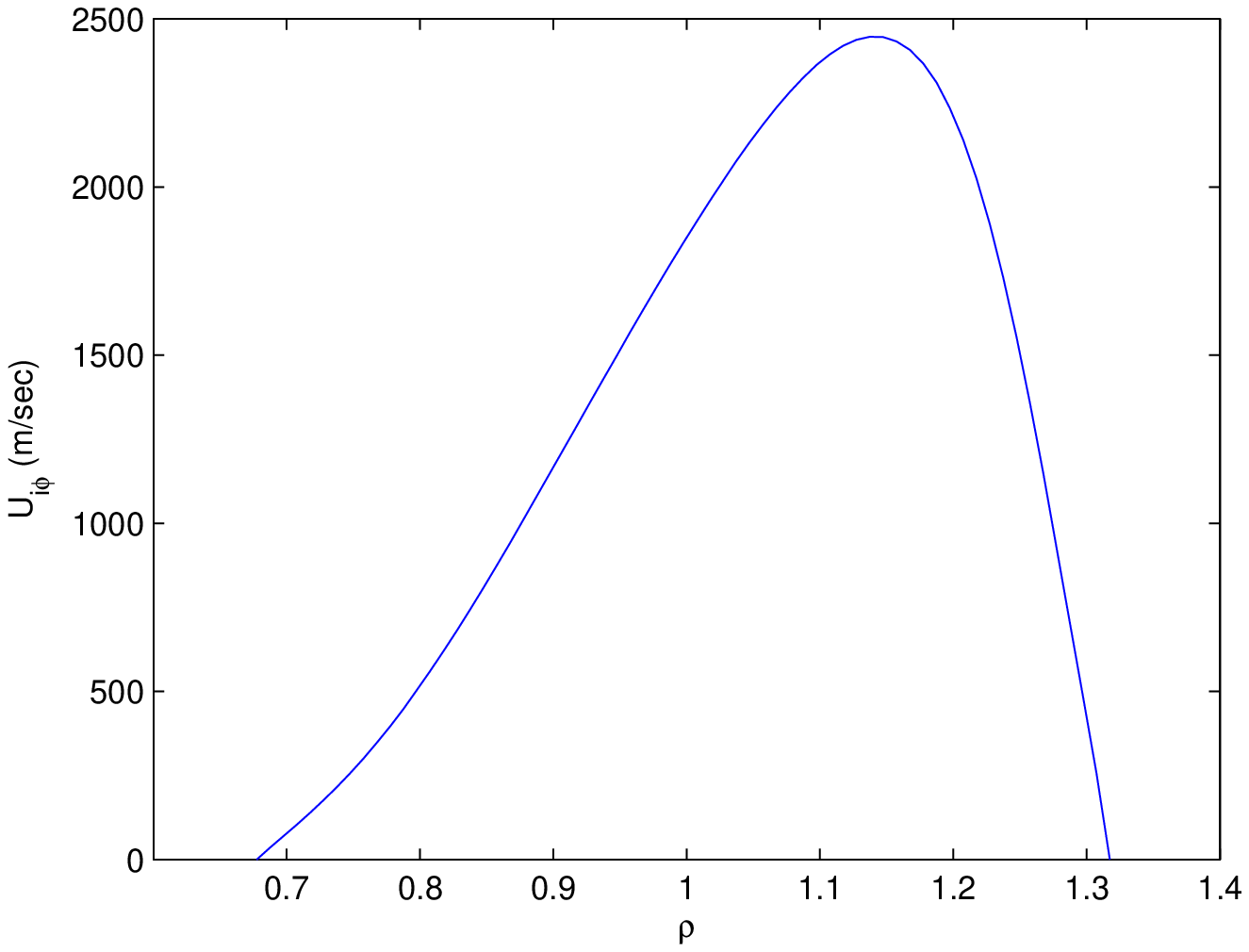}}}
\centerline{\mbox {\epsfxsize=10.cm \epsfysize=8.cm \epsfbox{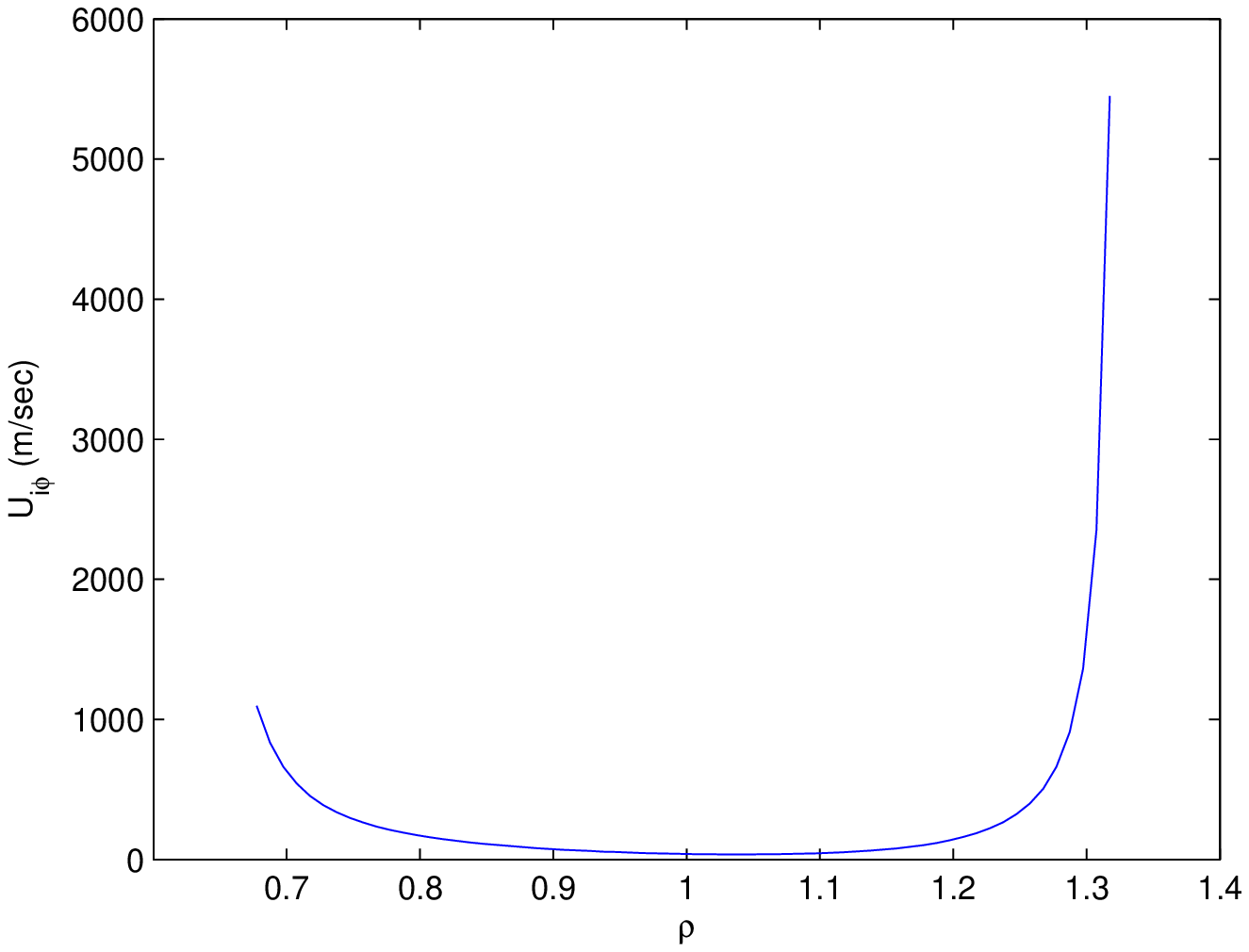}}}
\caption[]{The ion fluid velocity profile on the midplane $z=0$ for the equilibrium with peaked $J_\phi$ 
%(Fig. 5a) 
(above)
and hollow $J_\phi$ 
%(Fig. 5b). 
(below).}
\label{fig5}
\end{figure}
\begin{figure}[ht!]
\centerline{\mbox {\epsfxsize=10.cm \epsfysize=7.5cm \epsfbox{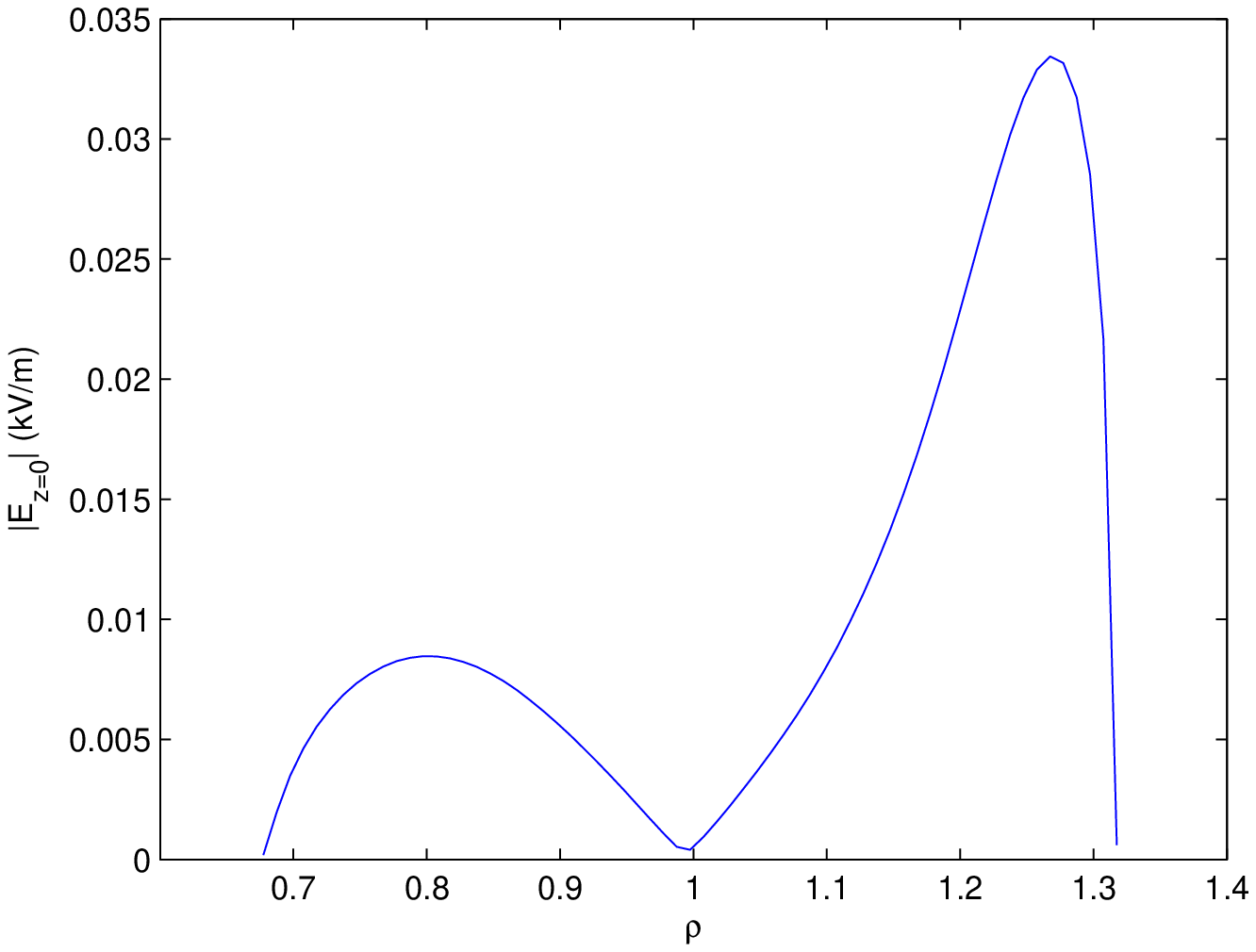}}}
\centerline{\mbox {\epsfxsize=10.cm \epsfysize=8.cm \epsfbox{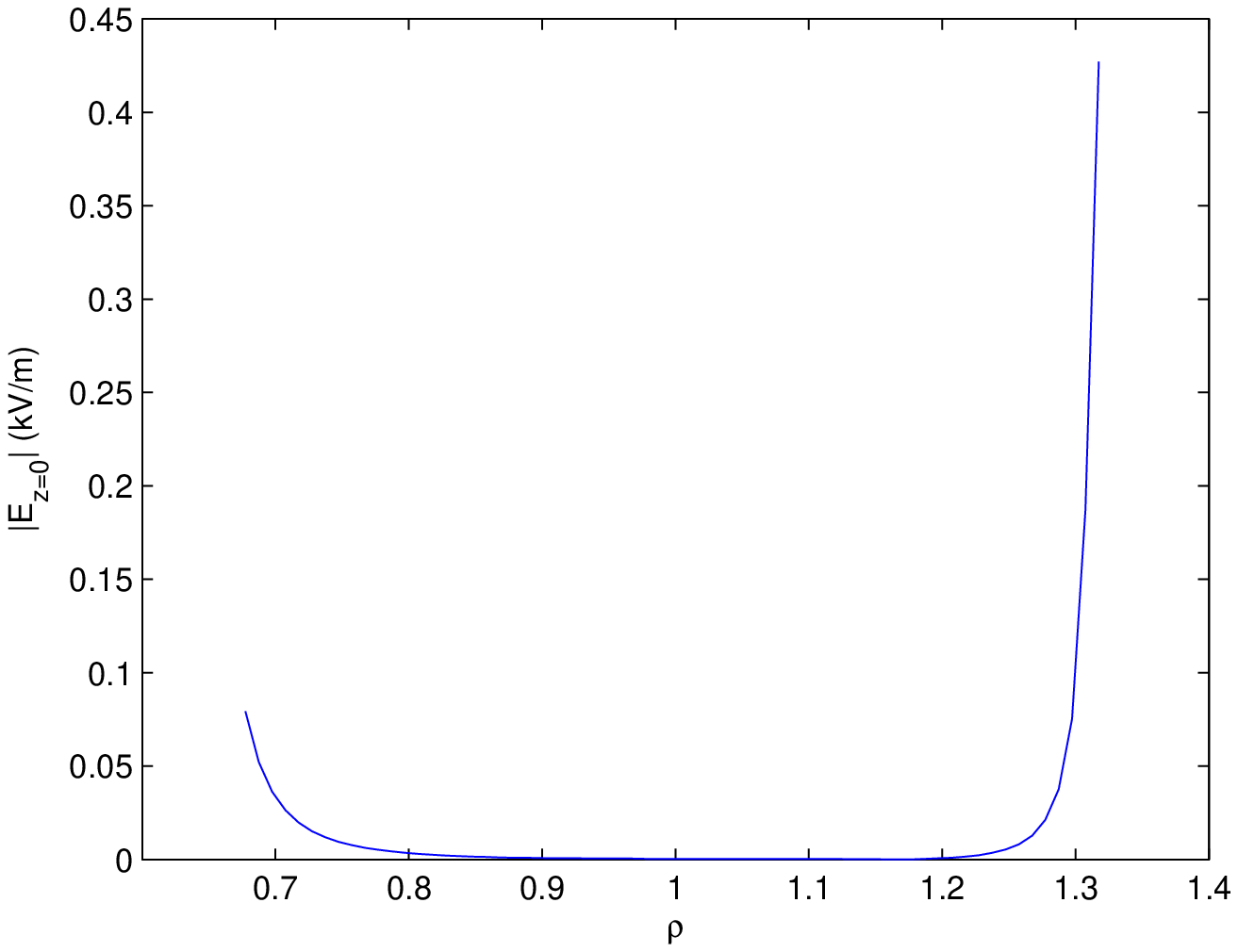}}}
\caption[]{The electric field  profile on the midplane $z=0$ for the equilibrium with peaked $J_\phi$ 
%(Fig. 6a)
(above)
and hollow $J_\phi$
% (Fig. 6b).
(below). }
\label{fig6}
\end{figure}

According to  \cite{frei} (p. 14 therein) the total pressure tensor
can be defined as
\bea
\label{presstens}
P_{kl}&=&P_{kl}^{i}+P_{kl}^{e}=\nn \\
&=&M\int(v_{k}-U_{ik})(v_{l}-U_{il})d^{3}v+\nn \\
&+&m\int(v_{k}-U_{ek})(v_{l}-U_{el})d^{3}v,\; \; \; (k,l=r,\phi,z) \nn
\\
\eea
where $P_{kl}^{i}$, $P_{kl}^{e}$ are the pressure tensors of the ion and
electron fluids and $U_{i\phi},\; U_{e\phi}$ are the ion and electron
fluid velocities given by $U_{i\phi}=(1/n_{i})\int v_{\phi}f_{i}d^{3}v$
and $U_{e\phi}=(1/n_{e})\int v_{\phi}f_{e}d^{3}v$. In the limit of
$V_{i\phi}\longrightarrow 0$ we recover the usual formula of
$n_{0}k(T_{i}+T_{e})$ for the total pressure. For the distribution functions
(\ref{distrs}) the tensor becomes diagonal and anisotropic and we have
$P_{rr}=P_{zz}:=P_{1}$ and $P_{\phi\phi}:=P_{2}$. Here
\bea
\label{p1}
P_{1}&:=&MI_{i}+mI_{e}=M\int v_{r}^{2}f_{i}d^{3}v+m\int v_{r}^{2}f_{e}d^{3}v\nn
\\
P_{2}&=&MJ_{i}+mJ_{e}=M\int(v_{\phi}-U_{i\phi})^{2}f_{i}d^{3}v+\nn
\\
&+&m\int(v_{\phi}-U_{e\phi})^{2}f_{e}d^{3}v\nn \\
\eea
Computing the integrals $I_{i},\; I_{e},\; J_{i},\; J_{e}$ we calcualted $P_1$, $P_2$ and the 
index of anisotropy
\bea
\label{deiktns}
\omega:=\frac{|P_{2}-P_{1}|}{P_{1}}
\eea
The calculation though  lengthy and tedious is straightforward. According to the  results shown in Fig. 7, $P_1$ and $P_2$ decrease very weakly from the magnetic axis to the plasma boundary, irrespective of the shape of the current density profile.Similar is the behavior of the density variation given
in Fig. 8 but the anisotropy index shown in Fig. 9 is different; the latter follows the respective variation of the current 
density profile.
Apparently, such nearly flat pressure and density profiles are not representative  for tokamaks. They just may be regarded  as an  approximation of the nearly flat pressure and density profiles observed during the L-H transition in the central region inside  the edge pedestal at which  the pressure drops  sharply giving rise to the ELMs activity.  This edge region can not be described  by the equilibrium solutions  constructed  here; in particular,   we had a   difficulty  to make $P_1$ and $P_2$  vanish on the boundary. This in addition to numerical reasons should be related with the fact that  Vlasov theory involves  particle orbits which, unlike fluid theories,  are not compatible with a fixed boundary; in this sense Vlasov  is not appropriate to describe fixed boundary equilibria in the  region  close to the boundary.
\begin{figure}[ht!]
\centerline{\mbox {\epsfxsize=10.cm \epsfysize=8.cm \epsfbox{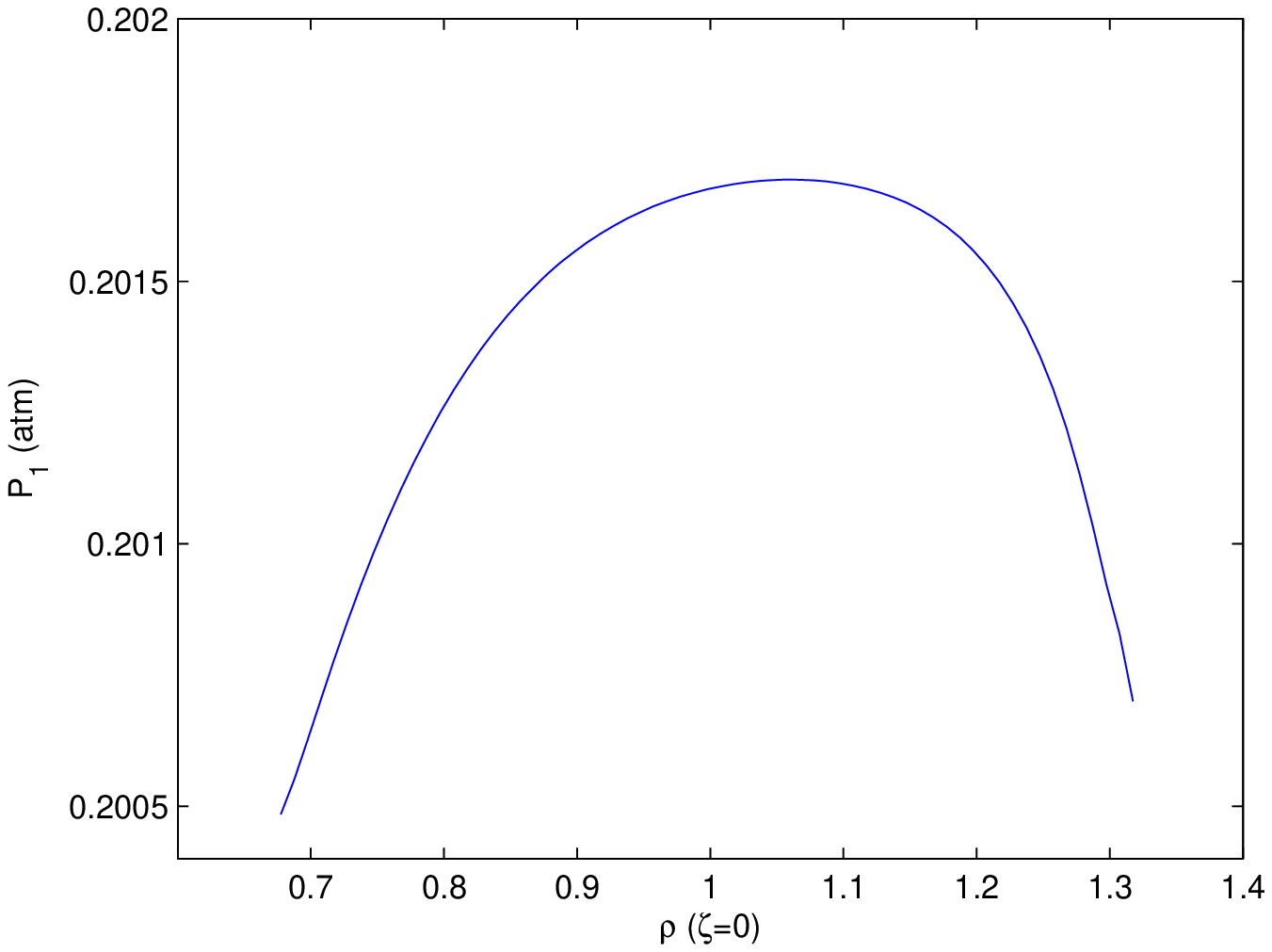}}}
\centerline{\mbox {\epsfxsize=10.cm \epsfysize=8.cm \epsfbox{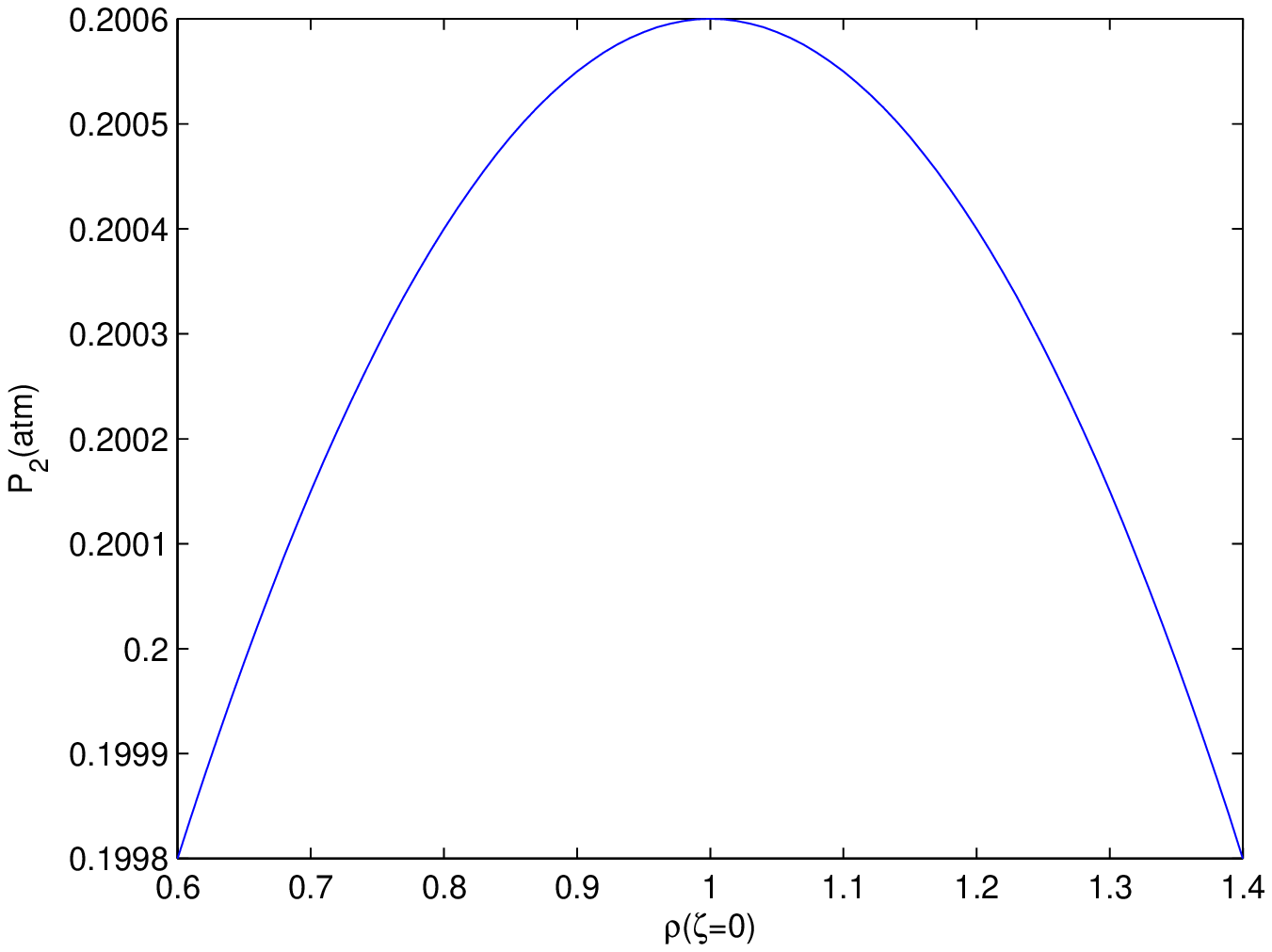}}}
\caption[]{Profiles of the pressures  $P_1$ and $P_2$ for the equilibrium with peaked $J_\phi$. Similar in shape   are  the respective pressure profiles with hollow $J_\phi$. }
\label{fig7}
\end{figure}
\begin{figure}[ht!]
\centerline{\mbox {\epsfxsize=10.cm \epsfysize=8.cm \epsfbox{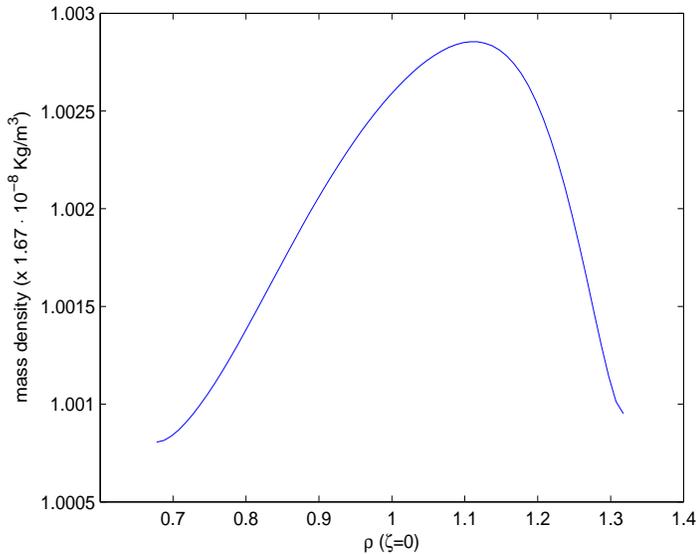}}}
\caption[]{Density profile for the equilibrium with peaked $J_\phi$. Similar in shape   is the density profile  for the equilibrium with hollow $J_\phi$.  }
\label{fig8}
\end{figure}
\begin{figure}[ht!]
\centerline{\mbox {\epsfxsize=10.cm \epsfysize=8.cm \epsfbox{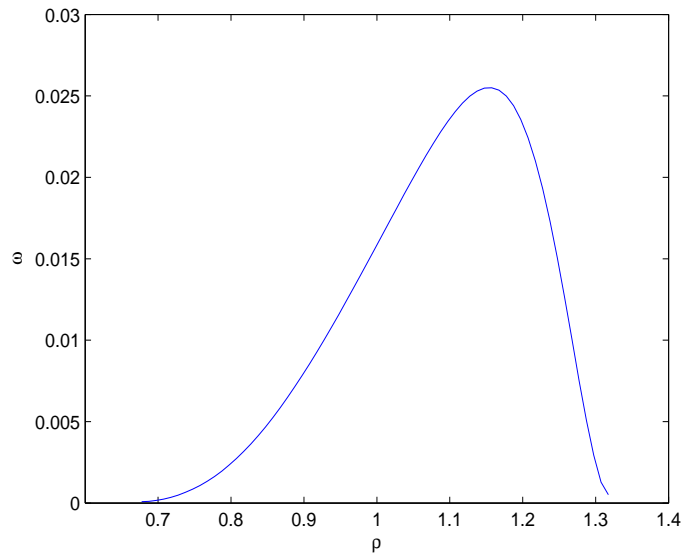}}}
\centerline{\mbox {\epsfxsize=10.cm \epsfysize=8.cm \epsfbox{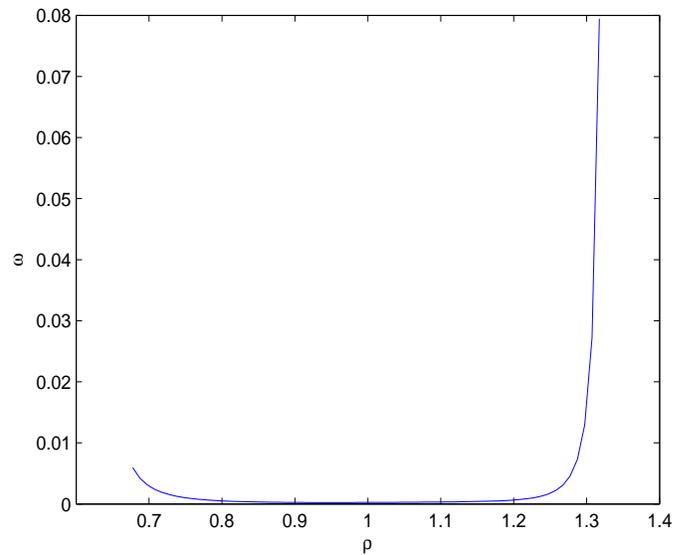}}}
\caption[]{The pressure anisotropy index for the equilibrium with peaked $J_\phi$ 
%(Fig. 9a)
(above)
and hollow $J_\phi$ 
%(Fig. 9b)
(below).}
\label{fig8}
\end{figure}

\section{Summary}

In the framework of Maxwell-Vlasov theory we have derived a  transcendental GS-like equation [Eqs. (17)-(18)]  for quasineutral axisymmetric plasmas by selecting  exponentially deformed Maxwellian ion and electron distribution functions depending on the two known constants of particle motion [Eqs. (10)]. Then we have solved this  equation  numerically  for a tokamak pertinent fixed diverted plasma boundary. To avoid unphysical perpendicular drifts on axis the electric filed was provisioned to vanish thereon. Depending on the symmetry properties of the distribution functions in connection with the values of pertinent free parameters,  we derived equlibria  with  toroidal current density profiles either peaked on axis or hollow. Both equilibria have nearly the same magnetic surfaces,
% safety factors  increasing monotonically from the magnetic axis to the plasma boundary,  
sheared toroidal ion flow and diagonal anisotropic pressure tensor  with  different toroidal and poloidal elements.   The profiles of the toroidal velocity and of the pressure isotropy index are similar to the respective profiles of the current density. However, the profiles of the pressure elements  being nearly flat just may approximately represent  respective experimental profiles in the central region developed during the L-H transition.  

It is interesting to pursue obtaining other Vlasov equilibria with alternative choices of distribution functions, in particular distribution functions potentially creating more realistic pressure profiles. This may require numerical integrations in the velocity space which  coupled with  spatial integrations constitute a challenging problem.  Also,  since poloidal flows play a role in the transitions to improved confinement regimes in tokamaks it is  desirable that these  equilibria involve flows  of arbitrary direction. However, as already mentioned in Sec.  I, this remains a tough open problem requiring  additional Vlasov constants of motion.

%%%%%%%%%%%%%%%%%%%%%%%%%%%%%%%%%%%%%%%%%%%%%%%%%%%%%%%%%%%%%%%%%%%%%
\vspace{1cm}

\section*{Aknowledgment}\

%One of the authors (GNT) would like to thank Drs. Henri Tasso and
%Calin Atanasiu for very useful discussions.

This work has been carried out within the framework of the EUROfusion Consortium and has received funding from  (a) the National Programme for the Controlled Thermonuclear Fusion, Hellenic Republic, (b) Euratom research and training programme 2014-2018 under grant agreement No 633053. The views and opinions expressed herein do not necessarily reflect those of the European Commission.

%%%%%%%%%%%%%%%%%%%%%%%%%%%%%%%%%%%%
%\Sec. *{APPENDIX}

%%%%%%%%%%%%%%%%%%%%%%%%%%%%%%%%%%
%%%%%%%%%%%%%%%%%%%%%%%%%%%%%%%%%%%%%%%%%%%%%%%%%%%%%%%%%%%%%%%%%%%%%%%%%%%%%%%%%%%%%%%%%%%%%%%%%
\newpage

\bibliography{apssamp}% Produces the bibliography via BibTeX.

\end{document}